\documentclass[aps,pre,preprint,groupedaddress,showkeys]{revtex4-1}

\usepackage{graphicx}
\usepackage{amssymb,amsfonts,amsmath}
\usepackage{fullpage}
\usepackage{longtable}
\usepackage{xcolor}

\begin{document}


\title{Deionization Shocks in Crossflow}


\author{Sven Schlumpberger}
\affiliation{Department of Chemical Engineering, Massachusetts Institute of Technology, Cambridge, MA 02139}
\author{Raymond B. Smith}
\affiliation{Department of Chemical Engineering, Massachusetts Institute of Technology, Cambridge, MA 02139}
\author{Huanhuan Tian}
\affiliation{Department of Chemical Engineering, Massachusetts Institute of Technology, Cambridge, MA 02139}
\author{Ali Mani}
\affiliation{Department of Chemical Engineering, Massachusetts Institute of Technology, Cambridge, MA 02139}
\author{Martin Z. Bazant}
\email[]{bazant@mit.edu}
\affiliation{Department of Chemical Engineering and Department of Mathematics, Massachusetts Institute of Technology, Cambridge, MA 02139}



\begin{abstract}
Shock electrodialysis is a recently developed electrochemical water treatment method which shows promise for water deionization and ionic separations. Although simple models and scaling laws have been proposed, a predictive theory has not yet emerged to fit experimental data and enable system design. 
Here, we extend and analyze existing "leaky membrane" models for the canonical case of a steady shock in cross flow, as in recent experimental prototypes. Two-dimensional numerical solutions are compared with analytical boundary-layer approximations and experimental data. The boundary-layer theory accurately reproduces the simulation results for desalination, and both models predict the data collapse of the desalination factor with dimensionless current, scaled to the incoming convective flux of cations. The numerical simulation also predicts the water recovery increase with current. Nevertheless, both approaches cannot quantitatively fit the transition from normal to over-limiting current, which suggests gaps in our understanding of extreme electrokinetic phenomena in porous media. 
\end{abstract}


\maketitle

\section{Introduction}
\label{intro}

The lack of access to clean water is a global public health challenge. According to the World Health Organization, only seven out of ten people used safely managed water in 2015 \cite{WorldHealthOrganizationWHOtheUnitedNationsChildrensFundUNICEF2017Progress2017}, which highlights the importance of developing or improving technologies for desalination, decontamination, and disinfection of water \cite{shannon_science_2008}. Shock electrodialysis (ED) is a recently proposed water treatment technology~\cite{wenten2020novel}, based on the mechanism of deionization shock waves in charged porous media~\cite{mani_deionization_2011} and microchannels~\cite{mani_propagation_2009,yaroshchuk_over-limiting_2012}, and primary results have shown its capability for deionization, filtration, separation, and disinfection \cite{deng_overlimiting_2013, deng_water_2015, Schlumpberger_Scalable_2015}. Compared with traditional technologies like reverse osmosis, thermal distillation, and electrodialysis, shock electrodialysis is scalable, continuous, potentially membrane free, and efficient at low ionic strength~\cite{wenten2020novel}. 

\begin{figure}[h!]
    \centering
    \includegraphics[width = 0.7 \textwidth]{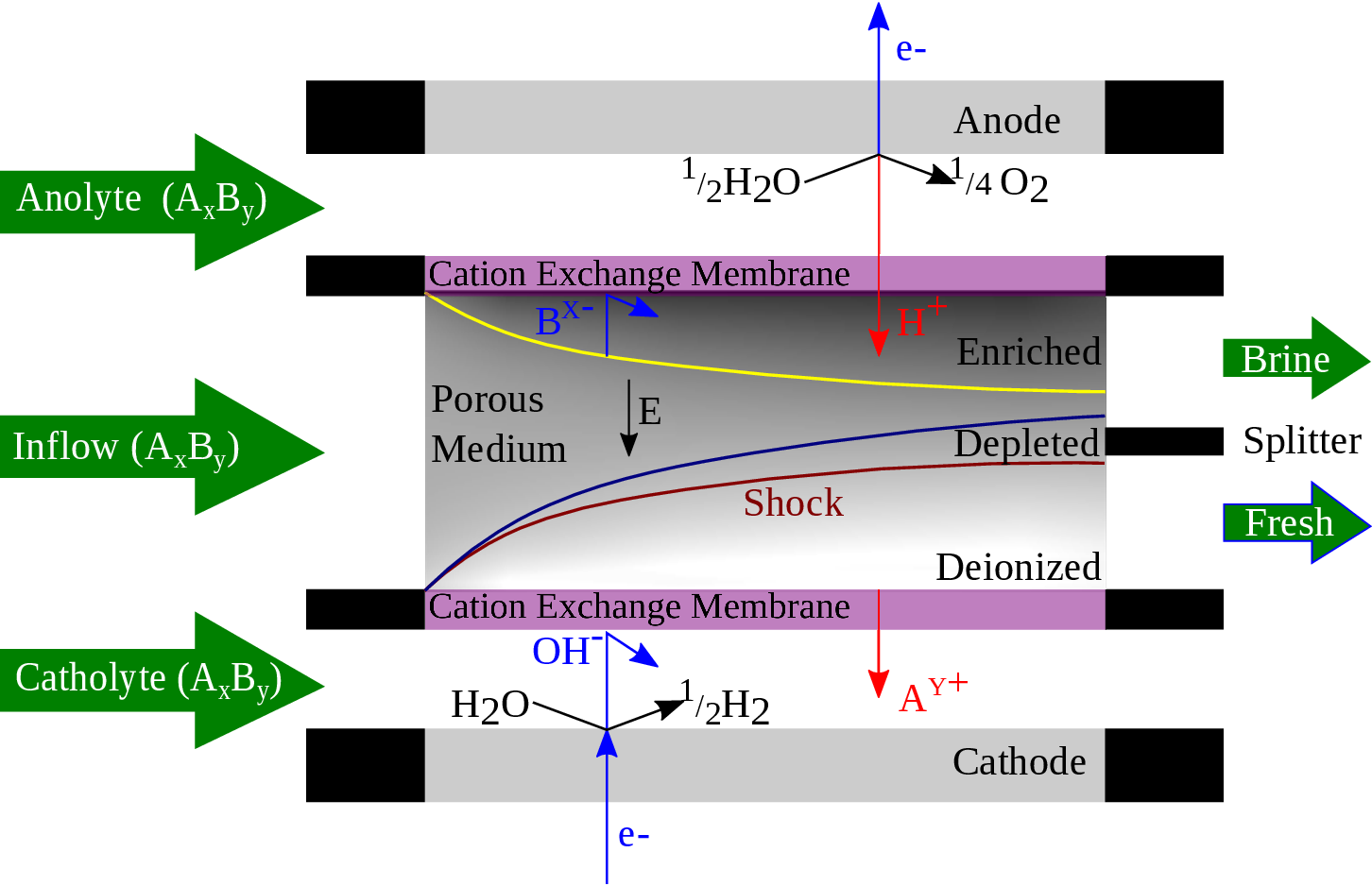}
    \caption{A representation of the shock electrodialysis prototype from Ref. \cite{Schlumpberger_Scalable_2015}, as first suggested in Ref.\cite{deng_overlimiting_2013}. }
    \label{fig:prototype}
\end{figure}

The first shock ED prototype for scalable, continuous operation\cite{Schlumpberger_Scalable_2015} employs a cross-flow geometry sketched in Figure \ref{fig:prototype}, where a charged glass frit is sandwiched between cation exchange membranes and electrode streams. The original study demonstrated desalination of simple salt solutions by 99\% at 65\% water recovery, with the possibility of up to 99.99\% desalination and 79\% water recovery~\cite{Schlumpberger_Scalable_2015}.  Similar designs have been employed in recent work demonstrating multivalent ionic separation~\cite{Conforti2019ContinuousElectrodialysisb}, nuclear wastewater treatment\cite{Alkhadra2019ContinuousElectrodialysis}, and small-scale seawater desalination~\cite{Alkhadra2020Small-scaleElectrodialysis}. Understanding which quantities govern the formation and propagation of deionization shocks in these situations would be vital for optimizing the prototype and developing new designs for specific applications. 

Shock electrodialysis relies on the phenomenon of overlimiting current in electrochemical systems \cite{dydek_overlimiting_2011, nam2015}. Consider the prototype shown in Figure \ref{fig:prototype}. Two cation-exchange membranes are in contact with a porous medium. Once an electrical field is applied across the membranes, cations are driven out near the cathode, while anions near the anode are repelled back by the membrane. Therefore, ions get depleted near the cathode while accumulating near the anode, which is usually referred to as concentration polarization \cite{zangle_propagation_2009, nielsen_concentration_2014}. If the electrical field is strong enough, theoretically, the ions near the cathode-side membrane are completely depleted, which leads to loss of conductance and a current limit \cite{dydek_overlimiting_2011}. However, in practice, overlimiting current is usually observed, which may be attributed to surface conduction \cite{dydek_overlimiting_2011}, electro-osmotic flow \cite{yaroshchuk_coupled_2011,dydek_overlimiting_2011}, or non-ideal selectivity of the membrane \cite{andersen_current-induced_2012}. Specifically, in the prototype of Ref. \cite{Schlumpberger_Scalable_2015}, the porous medium is an ultrafine frit (Adam \& Chittenden Scientific Glass), with a pore size of 0.9 - 1.4 ${\mu \mathrm{m}}$ and negative surface charge in contact with the electrolyte. In this case, surface conductance should be the dominant mechanism in the depleted region \cite{dydek_overlimiting_2011}. In other words, though the anions are almost completely depleted, some cations still remain to compensate for the surface charge and carry the current. Therefore, by increasing the electrical field, the depleted region can propagate from the cathode-side membrane into the frit. By imposing a cross-flow and setting a splitter near the middle of the outlet, the device can produce a desalinated and a brine stream. The transient propagating front of the deionized region between infinitely distant membranes is similar to gas shocks  \cite{mani_propagation_2009, mani_deionization_2011}, which is how shock electrolysis gets its name.   

Recently, models that aim to analyze the desalination shock phenomenon in various contexts have been developed. They first started with 1-D models in microfluidic channels \cite{zangle_propagation_2009, mani_propagation_2009, dydek_overlimiting_2011} that were then extended to porous media \cite{mani_deionization_2011, deng_overlimiting_2013} and multiple dimensions \cite{dydek_nonlinear_2013}. For porous media in particular, the 1-D models have been fairly successful in providing scaling estimates for quantities such as the overlimiting conductance for porous media \cite{deng_overlimiting_2013}, despite containing numerous simplifying assumptions and neglecting effects, such as dispersion and electro-osmotic flow. However, the 1-D model is not sufficient for our shock electrodialysis prototype due to the cross flow. A two-dimensional model has also been proposed \cite{dydek_nonlinear_2013}, which assumes both a symmetric electrolyte (binary electrolyte with the same valance and diffusivity for cation and anion) and simple plug flow through the porous medium. However, it is unable to explain the water recovery trends that were observed in experiments~\cite{Schlumpberger_Scalable_2015}. 

In this work, we first write down the general equations for shock ED with asymmetric electrolytes and electro-osmotic flow. Then we derive a simple analytical model based on boundary layer theory and an assumption of plug flow. Next, we present a simulation of the full model with incorporation of electro-osmotic flow in order to capture the flow that occurs from anode to cathode in experiment, which is critical for the prediction of water recovery. Finally, we  compare these theoretical predictions with experimental data from Ref. \cite{Schlumpberger_Scalable_2015} in order to test the underlying models. 

\section{Governing equations}
\label{govEq}
In this paper, we consider a homogeneous porous structure with finite surface charge density $\sigma_s$ (charge/pore area), area density $a_p$ (pore area/total volume), and porosity $\epsilon_p$ (pore volume/ total volume). The characteristic pore size is defined as $h_p = \epsilon_p/a_p$. We assume that we have only one simple electrolyte and ignore hydrogen and hydroxide ions. The species to be considered are water, cations of charge $z_+$, and anions of charge $z_-$.


\begin{center}
\begin{longtable}{ |c|c|c|c| } 
\caption{Parameters and Variables for the Shock Electrodialysis Prototype.}\\
\hline
\bf Parameter & \bf Symbol & \bf Value & \bf Dimensionless Scale\\
\hline
\endfirsthead
\multicolumn{4}{c}{\tablename\ \thetable\ -- \textit{continued from previous page}}\\
\hline
\bf Parameter & \bf Symbol & \bf Value & \bf Dimensionless Scale\\
\hline
\endhead
\hline
\multicolumn{4}{r}{\textit{Continued on next page}}\\
\endfoot
\hline
\endlastfoot
Boltzmann Constant & $k$ & $1.381\times 10^{-23}$ $\frac{J}{K}$ & -- \\ 
Absolute Temperature & $T$ & 293 K & -- \\ 
Electronic Charge & $e$ & $1.602\times 10^{-19}$ C & -- \\ 
Viscosity of Water & $\mu$ & 0.001002 Pa*s & -- \\
Permittivity of Water & $\epsilon_w$ & $6.947\times 10^{-10}$ $\frac{F}{m}$ & -- \\
Silanol Surface Density & $\Gamma$ & $8\times 10^{-18}$ m$^{-2}$ & --\\
Silanol pK & $pK$ & 7.5 & --\\
Solution pH & $pH$ & 7.0 & --\\
Stern Layer Capacitance & $C$ & 2.9 $\frac{F}{m^2}$ & --\\
Porosity & $\epsilon_p$ & 0.48 & -- \\
Porous Structure Area Density & $a_p$ & $1.785\times 10^{6}$ $m^{-1}$ & -- \\
Characteristic Pore Size & $h_p$ & $\epsilon_p/a_p$ ($2.689\times 10^{-7}$ m) & -- \\
Cation Charge Number & $z_+$ & 1 & -- \\
Anion Charge Number & $z_-$ & -1 or -2 & -- \\
Cation Stoichiometric Coeff. & $s_+$ & 1 or 2 & -- \\
Anion Stoichiometric Coeff. & $s_-$ & 1 & -- \\
Inlet Concentration & $c_0$ & Varies (m$^{-3}$) & -- \\
Outlet Pressure & $P_1$ & 10$^5$ Pa & -- \\
Inlet Pressure & $P_0$ & Varies & $P_1$ \\
Height of Porous Medium & $H$ & 2.7 mm & $H$ \\
Length of Porous Medium & $L$ & 10 mm & $H$ \\
Width of Porous Medium & $W$ & 20 mm & $H$ \\
Cation Diffusion Coefficient & $D_+$ & Varies $\left(\frac{m^2}{s}\right)$ & $D_-$ \\
Anion Diffusion Coefficient & $D_-$ & Varies $\left(\frac{m^2}{s}\right)$ & -- \\
Electro-osmotic Permeability & $k_{EO}$ & Varies = $-\frac{\epsilon_w\zeta}{\mu}$ $\left(\frac{m^2}{V*s}\right)$ & $\frac{D_-e}{kT}$ \\
Darcy Permeability & $k_D$ & $h_p^2/2\mu$ ($3.608\times 10^{-11}$ $\frac{m^2}{Pa*s}$) & $\frac{D_-}{P_1}$ \\
Surface Charge & $\sigma_s$ & Varies $\left(\frac{C}{m^2}\right)$ & -- \\
Volumetric Surface Charge & $\rho_s$ & Varies = $\sigma_s/h_p$ $\left(\frac{C}{m^3}\right)$ & Sim: $es_-c_0$; BL: $ec_{b,0}$ \\
Effective Volumetric Charge & $\rho_{eff}$ & Varies = $\frac{k_{EO}}{k_D}$ $\left(\frac{C}{m^3}\right)$ & $\frac{eP_1}{kT}$\\
Applied Current & $I_{app}$ & Varies (A) & $D_-s_-c_0eH$ \\
Applied Voltage & $V_{app}$ & Varies (V) & $\frac{kT}{e}$ \\
Zeta Potential & $\zeta$ & Varies (V) & --\\
Potential Perturbation & $\psi$ & Varies (V) & --\\
\hline
\bf Variable & \bf Symbol & \bf Value & \bf Dimensionless Scale\\
\hline
Ionic Potential & $\phi$ & -- & $\frac{kT}{e}$ \\
Anion Concentration & $c_-$ & -- & $s_-c_0$ \\
Cation Concentration & $c_+$ & -- & $s_-c_0$ \\
Neutral Bulk Concentration & $c_b$ & $-2z_-c_-$ & $c_{b,0}$ \\
Pressure & $P$ & -- & $P_1$ \\
Position & ($x$,$y$) & -- & ($H$,$H$) \\
Anion Flux & \textbf{F}$_-$ & -- & $\frac{D_-s_-c_0}{H}$ \\
Cation Flux & \textbf{F}$_+$ & -- & $\frac{D_-s_-c_0}{H}$ \\
Flow Velocity & \textbf{u} & -- & $\frac{D_-}{H}$ \\
Current Density & \textbf{J} & -- & $\frac{D_-s_-c_0e}{H}$
\label{paramTab}
\end{longtable}
\end{center}

First, the species flux, $\textbf{F}_i$, and the ionic current density, $\textbf{J}$, can be expressed using the Nernst-Planck equation, assuming dilute solution theory:
\begin{equation}
\textbf{F}_- = \epsilon_p^{3/2}\left(-D_-\nabla c_- + \textbf{u}c_- - \frac{D_-z_-e}{kT}c_-\nabla\phi\right),
\label{fluxn}
\end{equation}
\begin{equation}
\textbf{F}_+ = \epsilon_p^{3/2}\left(-D_+\nabla c_+ + \textbf{u}c_+ - \frac{D_+z_+e}{kT}c_+\nabla\phi\right),
\label{fluxp}
\end{equation}
\begin{equation}
\textbf{J} = z_-e\textbf{F}_- + z_+e\textbf{F}_+.
\label{currentd}
\end{equation}
Furthermore, the flow velocity in the pores, \textbf{u}, can generally be expressed using a combination of Darcy's law and linear electro-osmosis:
\begin{equation}
\textbf{u} = -k_{EO}\nabla\phi - k_D\nabla P,
\label{vel}
\end{equation}
where the first term is electro-osmotic flow and the second is Darcy's law. $\epsilon_p^{3/2} \mathbf{u}$ would be the traditional superficial velocity in porous media. For a pore with circular cross section $k_D=h_p^2/2\mu$, where $\mu$ is the electrolyte viscosity. $k_{EO}$ can be approximated by the Helmholtz-Smoluchowski equation $k_{EO}=-\frac{\epsilon_w\zeta}{\mu}$,
where $\epsilon_w$ is the electrolyte permittivity, and $\zeta$ is the surface potential, which would be negative for a negatively charged solid phase. $\zeta$ can be nonuniform due to variation of concentration \cite{gentil_transistor_2006}. In this work, for simplicity, we assume a uniform $\zeta$. 

The conservation of ionic species and incompressibility then give us the following governing equations:
\begin{equation}
\frac{\partial (\epsilon_p c_-)}{\partial t} + \nabla\cdot\textbf{F}_- = 0,
\label{massconsn}
\end{equation}
\begin{equation}
\frac{\partial (\epsilon_p c_+)}{\partial t} +\nabla\cdot\textbf{F}_+ = 0,
\label{massconsp}
\end{equation}
\begin{equation}
\nabla\cdot\textbf{u} = 0.
\label{incomp}
\end{equation}

Finally, we assume electroneutrality in the charged porous medium: 
\begin{equation}
z_+ec_+ + z_-ec_- + \rho_s = 0.
\label{eneut}
\end{equation}
where $\rho_s=\frac{\sigma_s}{h_p}$ is the pore-volume-averaged surface charge density. In this model, we assume constant $\rho_s$.

\section{Boundary layer analysis}

First, we look at deriving a simple solution based on boundary layer theory. Without imposed flow, the system would reach a one-dimensional concentration profile (away from the edges) with concentration linearly dependent on the wall-normal direction \cite{dydek_overlimiting_2011}. However, with flow present (we assume plug flow in this part), the competition between advection, which brings fresh salt into the system, and concentration polarization, which acts to deplete the salt next to the membrane, leads to the formation of a diffusion boundary layer structure growing next to the membrane \cite{dydek_nonlinear_2013}. 
This boundary layer consists of two regions: 
\begin{enumerate}
\item The outer (non-depleted) region is a diffusion boundary layer with characteristics similar to that in an unsupported electrolyte.
\item The inner (depleted) region consists of almost desalted bulk fluid. In this region, surface conduction dominates electric current.
\end{enumerate}
The two regions are separated by a propagating jump in salt concentration shock, referred to as a ``deionization shock" \cite{mani_deionization_2011}. In the following, we will adopt a boundary layer analysis for a system involving porous media to obtain a matched asymptotic approximation for nested boundary layers, which can be used to predict deionization by shock ED.

\subsection{Assumptions and Equations}
\label{governingequations}
We consider a two-dimensional domain describing flow of a salty electrolyte in a frit over a single membrane. The {\it x}-direction represents the streamwise direction parallel to the membrane and the {\it y}-direction is normal to the membrane. 
Here we use the equivalent expression of Eq.\ref{massconsn}-\ref{massconsp} at steady state \cite{mani_deionization_2011}
\begin{equation}
    \nabla \cdot \mathbf{J} = 0, 
\end{equation}
\begin{equation}
\label{eq:transport}
{\bf u \cdot \nabla}c_b=\bar{D}\left[ \nabla^2c_b-{\bf \nabla \cdot }\left( \frac{\bar{z}\rho_s}{e}{\bf \nabla}\tilde{\phi}\right)\right],
\end{equation} 
where $c_b=-2z_-c_-$ (number/pore volume) is the {\it neutral} bulk concentration summing cations and anions but excluding the excess counterions which balance the surface charge in the EDLs,  $\bar{z} = \frac{2z_-z_+}{z_--z_+}$, $\bar{D}=\frac{(z_+-z_-)D_+D_-}{z_+D_+-z_-D_-}$ is the ambipolar diffusivity, and $\tilde{\phi}=e\phi/kT$ is the nondimensionalized electrostatic potential. If $\rho_s \neq 0$, the ion transport will be coupled with the electric field, which leads to nonlinearity of the problem. 

For this analysis, we ignore electro-osmotic flow and assume the flow to be plug-flow in the $x$-direction:
\begin{equation}
    {\bf u} = U_0\hat{\bf x} = -k_D\frac{\partial P}{\partial x} \hat{\bf x},
    \label{eq:velBL}
\end{equation}
where $\hat{\bf x}$ is the unit vector in the $x$-direction. Combining Eqs. \ref{eq:transport} and \ref{eq:velBL}, we further simplify the equations by assuming that the electric field only acts in the $y$-direction and we make the standard boundary layer analysis assumptions that convection is dominant in the $x$-direction and diffusion is dominant in the $y$-direction. These assumptions yield

\begin{equation}
\label{eq:BL}
U_0\frac{\partial c_b}{\partial x}=\bar{D}\left[ \frac{\partial ^2 c_b}{\partial y^2}-\frac{\partial}{\partial y}\left( \frac{\bar{z}\rho_s}{e}\frac{\partial \tilde{\phi}}{\partial y}\right)\right],
\end{equation}
as well as conserved current density ${\bf J}$ in the $y$-direction:
\begin{equation}
    \mathbf{J} = \epsilon_p^{3/2}
    \left[ \left(\frac{1}{2}(z_+ D_+ - z_- D_-)e c_b + z_+D_+ \rho_s \right) \frac{ \partial \tilde{\phi}}{\partial y} + \frac{1}{2}e(D_+ - D_-) \frac{\partial c_b}{\partial y} \right] \hat{\bf y},
    \label{eq:ohm}
\end{equation}
where $\hat{\bf y}$ is the unit vector in the $y$-direction. The first term in the parentheses indicates the bulk conduction, the second indicates surface conduction carried by excess counterions in the EDLs, and the third indicates the diffusion current. 

Next we will apply Eq. \ref{eq:BL}-\ref{eq:ohm} to the inner depleted region and outer region with proper assumptions and boundary conditions, and get the asymptotic solution.

\subsection{Similarity solutions}

First, we assume that the depletion region (where $c_b \approx 0$) dominates the ohmic resistance along the almost vertical current lines. Given a depletion region with thickness $\delta_s(x)$, Eq. \ref{eq:ohm} can be simplified to:
\begin{equation}
\label{eq:normalcurrent}
{\bf J}=\epsilon_p^{3/2}\frac{V_{app}}{\delta_s}\nu_+ z_+ e \rho_s\hat{\bf y},
\end{equation}
where, $V_{app}$ is the applied voltage, $\nu_+ z_+ e \rho_s$ is the conductance in the depletion region solely provided by the counter-ions shielding the constant pore charge, and $\nu_+ = D_+/kT$ is the counterion mobility. 

In the non-depleted region of the boundary layer, we can further assume that the dynamics are governed by diffusion, and surface conduction plays a negligible role. Furthermore, we assume that the potential drop across this region is small, meaning that $\Delta c >> \frac{\rho_s}{e}\Delta \tilde{\phi}$. This allows us to reduce Eq.~\ref{eq:BL} in the outer/non-depleted region to 
\begin{equation}
\label{eq:BLsimpl}
U_0\frac{\partial c_b}{\partial x}=\bar{D}\frac{\partial ^2 c_b}{\partial y^2}.
\end{equation}
And the current density becomes
\begin{equation}
    \mathbf{J} = -\epsilon_p^{3/2} \frac{z_+ D_+ e}{\bar{z}} \frac{\partial c_b}{\partial y} \hat{\mathbf{y}},
    \label{eq:current_inner}
\end{equation}
which can be derived from the zero anion flux ($\frac{\partial c_-}{\partial y} + z_- c_- \frac{\partial \tilde{\phi}}{\partial y}=0$) and $J_y = z_+ e F_{y,+}$. Next we find a similarity solution to Eq. \ref{eq:BLsimpl} of the following form:
\begin{equation}
\label{eq:similarityassump}
c_b=c_{b,0}f(\eta),\ \eta=\frac{y}{2\sqrt{\bar{D}x/U_0}},
\end{equation}
where $c_{b,0}$ is the inlet bulk concentration of salt. Substitution of Eq.~\ref{eq:similarityassump} into the boundary layer Eq.~\ref{eq:BLsimpl} results in the following similarity equation for the outer region:
\begin{equation}
\label{eq:similarity}
-2\eta f^\prime = f^{\prime\prime}.
\end{equation}
Then we need to specify the boundary conditions. First, assume that the thickness of the depleted region $\delta_s$ is proportional to the thickness of the diffusion boundary layer $\delta = 2\sqrt{\bar{D} x/U_0}$ with a constant coefficient $\alpha$. In other words, $\eta |_{y=\delta_s}=\alpha$. 
Therefore, the dimensionless concentration $f$ needs to reach zero at the edge of the inner depleted region ($f(\alpha)=0$) and needs to reach 1 outside of the diffusion boundary layer ($f(\infty) = 1$). 
The solution of the above similarity equation predicts a concentration profile in the outer region:
\begin{equation}
\label{eq:outersolution}
f=1-\frac{\text{erfc}(\eta)}{\text{erfc}(\alpha)}.
\end{equation}
 We also define the 99\% diffusion boundary layer thickness as $\delta_d$, which satisfies $f(\beta) = 0.99$ with $\eta |_{y=\delta_d}= \beta$.
 
\begin{figure}
  \centering   
  \includegraphics[width=2.5in]{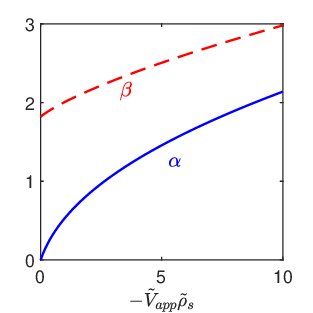}
  \caption{Dimensionless thickness of the depletion region, $\alpha=\frac{\delta_s}{2\sqrt{Dx/U_0}}$, and the dimensionless 99\% thickness of the diffusion boundary layer, $\beta = \frac{\delta_d}{2\sqrt{Dx/U_0}}$, as a function of $(-\tilde{V}_{app}\tilde{\rho_s})$ for a symmetric binary electrolyte.}
\label{Fig5}
\end{figure}

In order to determine $\alpha$, the dimensionless boundary layer thickness in units of the similarity variable, $\eta$, this solution must be matched to the inner solution. Equating the normal current density in the outer region (Eq. \ref{eq:current_inner})
and in the depleted region (Eq.~\ref{eq:normalcurrent}), we arrive at an algebriac relation for $\alpha$ 
\begin{equation}
\label{eq:alpha}
\frac{\alpha \exp(-\alpha^2)}{1-\text{erf}(\alpha)}= - \frac{\sqrt{\pi}}{2}  \bar{z} \tilde{V}_{app}\tilde{\rho_s}.
\end{equation}
where $\tilde{V}_{app} = V_{app} e/kT$ and $\tilde{\rho_s} = \rho_s/ec_{b,0}$. Figure~\ref{Fig5} shows a plot of this relation for a symmetric binary electrolyte, as well as the dimensionless 99\% boundary layer thickness $\beta$ as a comparison.

To make the model useful for comparison to experiment, we combine the outer solution from Eq. \ref{eq:outersolution} with the inner solution that simply assumes a fully depleted solution (i.e. $c_b = 0$) and obtain:
\begin{equation}
c_b = -2z_-c_- = c_{b,0}f(\eta) = \begin{cases} 0, & \mbox{if } \eta \leq \alpha \\ c_{b,0}\left(1-\frac{\text{erfc}(\eta)}{\text{erfc}(\alpha)}\right), & \mbox{if } \eta > \alpha \end{cases}
\end{equation}
The form of this model is the same as in analysis of reaction-diffusion fronts with neutral species, where the moving depleted region results from reactions of the diffusing reactant with a fixed reactant \cite{bazant_asymptotics_2000, leger_front_1999, koza_asymptotic_1997}. The main difference in this model as compared to prior analyses is that the depleted region has a residual concentration of cations that carry current by electromigration and also that the diffusivity in the diffusion layer is the ambipolar one. Furthermore, this model allows us to define a simple expression for total current:
\begin{equation}
\textbf{I} = I_{app} \hat{\bf y} = W\int_0^L \textbf{J} dx =
\epsilon_p^{3/2}\frac{V_{app}W\sqrt{U_0L}}{\alpha\sqrt{\bar{D}}}\nu_+ z_+e \rho_s\hat{\bf y}.
\label{BLcurr}
\end{equation}
where $W$ is the width of the porous medium. This equation can be combined with Eq. \ref{eq:alpha} to give the relation between $\alpha$ and $I$. The voltage across the porous layer is obtained from Eq.(\ref{eq:alpha}):
\begin{equation}
V_{app} = - \frac{2kTc_{b,0}}{\rho_s  \bar{z}\sqrt{\pi}}\frac{\alpha\exp{(-\alpha^2)}}{\text{erfc}(\alpha)},
\end{equation}
although this is difficult to compare with experimental data directly, due to voltage drops across the cation exchange membranes and the electrode interfaces~\cite{deng_overlimiting_2013}.
As we will see in a later section, this model offers some useful insight into the operation of the experimental system.

\section{Simulation}
For the simulation model, a more realistic geometry is used. The shock electrodialysis cell consists of two flat ion-exchange membranes with a charged porous medium between them, through which the electrolyte will flow. 
\subsection{Equations}
For this model, no further simplifying assumptions need to be made to the equations in section \ref{govEq} (Eqs. \ref{fluxn} -- \ref{eneut}). Recall that in section \ref{govEq} we have assumed constant $\rho_s$ and $\zeta$.
Now we can rearrange and combine these equations to give three governing equations of three variables of interest, which are the concentration of the anion $c_-$ (since the concentration of the cation is easily calculated from it through the electroneutrality condition), the electrical potential $\phi$, and the pressure $P$. 
The conservation of anion flux and ionic charge then lead to
\begin{equation}
\epsilon_p^{-1/2}\frac{\partial c_-}{\partial t} + \nabla\cdot\left[-D_-\left(\nabla c_- + \frac{z_-e}{kT}c_-\nabla\phi\right) - c_-\left(k_{EO}\nabla\phi + k_D\nabla P\right)\right] = 0,
\label{concEq}
\end{equation}
\begin{equation}
\nabla\cdot\left(k_{EO}\nabla\phi + k_D\nabla P\right) = 0,
\label{velEq}
\end{equation}
\begin{multline}
\nabla\cdot\left.\bigg\lbrace z_-e\left(D_+ - D_-\right)\nabla c_- + \frac{e}{kT}\left[\left(z_+D_+ - z_-D_-\right) z_-ec_- + D_+z_+\rho_s\right]\nabla\phi\right.\bigg\rbrace  = 0. 
\label{jEq}
\end{multline}

For ease of simulation, the dimensionless forms of these equations are used. The equations were non-dimensionalized using the dimensionless scales from Table \ref{paramTab}:

\begin{equation}
\epsilon_p^{-1/2}\frac{\partial \tilde{c}_-}{\partial\tilde{t}} + \tilde{\nabla}\cdot\left[-\left(\tilde{\nabla}\tilde{c}_- + z_-\tilde{c}_-\tilde{\nabla}\tilde{\phi}\right) - \tilde{c}_-\left(\tilde{k}_{EO}\tilde{\nabla}\tilde{\phi} + \tilde{k}_D\tilde{\nabla} \tilde{P}\right)\right] = 0,
\label{concEqNon}
\end{equation}
\begin{equation}
\tilde{\nabla}\cdot\left(\tilde{k}_{EO}\tilde{\nabla}\tilde{\phi} + \tilde{k}_D\tilde{\nabla} \tilde{P}\right) = 0,
\label{velEqNon}
\end{equation}
\begin{equation}
\tilde{\nabla}\cdot\left.\bigg\lbrace z_-\left(\tilde{D}_+ - 1\right)\tilde{\nabla} \tilde{c}_- + \left[\left(z_+\tilde{D}_+ - z_-\right) z_-\tilde{c}_- + \tilde{D}_+z_+\tilde{\rho}_s\right]\tilde{\nabla}\tilde{\phi}\right. 
\bigg\rbrace = 0.
\label{jEqNon}
\end{equation}

In addition, to calculate $\rho_s$ and $k_{EO}$, we need to respectively specify $\sigma_s$ and $\zeta$ for the silica frit, which are assumed to be dependent on inlet concentrations but not on current. We apply the following charge regulation model \cite{behrens_charge_2001, van_streaming_2005}:
\begin{equation}
\zeta (\sigma_s) = \frac{kT}{e}\left[\ln \left(\frac{-\sigma_s}{e\Gamma+\sigma_s}\right)+\ln (10)(pK-pH)\right] - \frac{\sigma_s}{C}
\label{zeta},
\end{equation}
\begin{equation}
\sigma_s (\zeta) = \text{sign}(\zeta) \sqrt{2kTc_0\epsilon_w\left[s_+\exp\left(-\frac{z_+e\psi}{kT}\right) + s_-\exp\left(-\frac{z_-e\psi}{kT}\right) - s_+ - s_-\right]}.
\label{sigma}
\end{equation}
The parameters we used for the model have been summarized in Table \ref{paramTab}, and the values for surface charge and zeta potential used in the simulations are shown in Table \ref{surfTab}.

\begin{table}
\begin{center}
\begin{tabular}{ |c|c|c| } 
\hline
Electrolyte Type / Concentration & Surface Charge (mC/m$^2$) & Zeta Potential (mV)\\
\hline
1:1 / 1 mM & -10.42 & -90.23\\ 
1:1 / 10 mM & -20.80 & -68.69\\ 
1:1 / 100 mM & -38.39 & -46.52\\ 
1:2 / 10 mM & -25.59 & -61.61\\
2:1 / 10 mM & -36.90 & -48.08\\
\hline
\end{tabular}
\caption{Surface charge and zeta potential for the shock electrodialysis prototype. The two numbers of the ratio in the first column represent the valence of the cation and anion of the electrolyte (i.e. $|z_+|:|z_-|$). }
\label{surfTab}
\end{center}
\end{table}
Lastly, we have to briefly consider two shortcomings of this model. First, it does not satisfy Onsager symmetry. This shortcoming stems from the volume averaging of the surface charge over the volume of the pore. In reality, the charge sits mostly in the electric double layers and hence experiences much less convection due to viscous drag than is predicted in this model. 
Furthermore, it is also important to note that this model ignores diffusioosmosis.  Second, we assume constant $\sigma_s$ and $\zeta$, which actually depend on local concentration and pH. We will consider these effects in future work. 

\subsection{Boundary Conditions}
\begin{figure}
\begin{center}
\includegraphics[width=10cm,keepaspectratio]{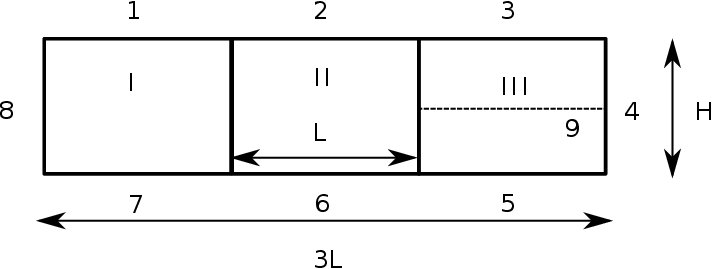}
\caption{Sketch of simulation domain.}
\label{sketch}
\end{center}
\end{figure}

The system has eight boundary segments (see Figure \ref{sketch}) for which we need to specify boundary conditions. Furthermore, the system has one internal boundary in domain III. We will only apply a potential or current in domain II. Domains I and III are intended to allow inlet and outlet flows to develop properly and to also facilitate application of inlet and outlet boundary conditions. See Table \ref{paramTab} for the values of $H$ and $L$. In addition, we place the splitter (boundary 9) at 0.454 of the channel height from boundary 5, so the water recovery at zero current should be 45.4\%, consistent with experimental data \cite{Schlumpberger_Scalable_2015}. 

\subsubsection{Wall Boundary Conditions (Boundaries 1, 3, 5, 7, \& 9)}
Boundaries 1, 3, 5, and 7 are solid outside walls and as such no fluxes are possible through these boundaries. Therefore, we have that $\hat{\bf n}\cdot\tilde{\textbf{F}}_- = \hat{\bf n}\cdot\tilde{\textbf{u}} = \hat{\bf n}\cdot\tilde{\textbf{J}} = 0$, where $\hat{\bf n}$ is a surface normal vector. In terms of the scalar variables used in Eqs. \ref{concEqNon} -- \ref{jEqNon}, these boundary conditions are $\hat{\bf n}\cdot\tilde{\nabla}\tilde{c}_-$ = $\hat{\bf n}\cdot\tilde{\nabla}\tilde{P}$ = $\hat{\bf n}\cdot\tilde{\nabla}\tilde{\phi}$ = 0. The same conditions are true at boundary 9, which is the flow splitter in the system and is modeled as an internal wall.

\subsubsection{Ion-Exchange Membranes (Boundaries 2 \& 6)}
Boundaries 2 and 6 are ion-exchange membranes. In this particular problem, we will only consider cation-exchange membranes, which means, assuming an ideal membrane, that the flux of anions through this boundary is zero ($\hat{\bf n}\cdot\tilde{\textbf{F}}_- = 0$). Furthermore, we will not allow any bulk flow through the membrane ($\hat{\bf n}\cdot\tilde{\textbf{u}} = 0$).
Lastly, we apply a uniform voltage on the membranes: $\tilde{\phi} = \tilde{V}_{app}/2$ on boundary 2 and $\tilde{\phi} = -\tilde{V}_{app}/2$ on boundary 6. The corresponding dimensionless applied current can be calculated from $\tilde{W}\int \tilde{\textbf{J}} \cdot \hat{\mathbf{n}} \,d\tilde{x}$ on either of boundary 2 or boundary 6. 

\subsubsection{Inlet Conditions (Boundary 8)}
Boundary 8 is the inlet to the system. Here we specify the inlet concentration ($\tilde{c}_- = 1$) and the inlet pressure ($\tilde{P} = \tilde{P}_0$). Furthermore, there should be no electric field in the normal direction at the inlet, so we impose $\frac{\partial \tilde{\phi}}{\partial \tilde{x}} = 0$.

\subsubsection{Outlet Conditions (Boundary 4)}
Boundary 4 is the outlet to the system. Here we only specify the outlet pressure ($\tilde{P} = 1$), which together with the inlet pressure condition determines the normal pressure drop across the system. Furthermore, since we are far removed from the region of applied current/potential, we can safely assume that there is no longer a normal electric field ($\frac{\partial \tilde{\phi}}{\partial \tilde{x}} = 0$). Lastly, we assume an outflow boundary condition for the anions, meaning we assume that the concentration no longer changes in the normal direction ($\frac{\partial \tilde{c}_-}{\partial \tilde{x}} = 0$).


\subsection{Simulation Results}
\begin{figure}
\begin{center}
\includegraphics[width=15cm,keepaspectratio,trim=0 90 0 80,clip]{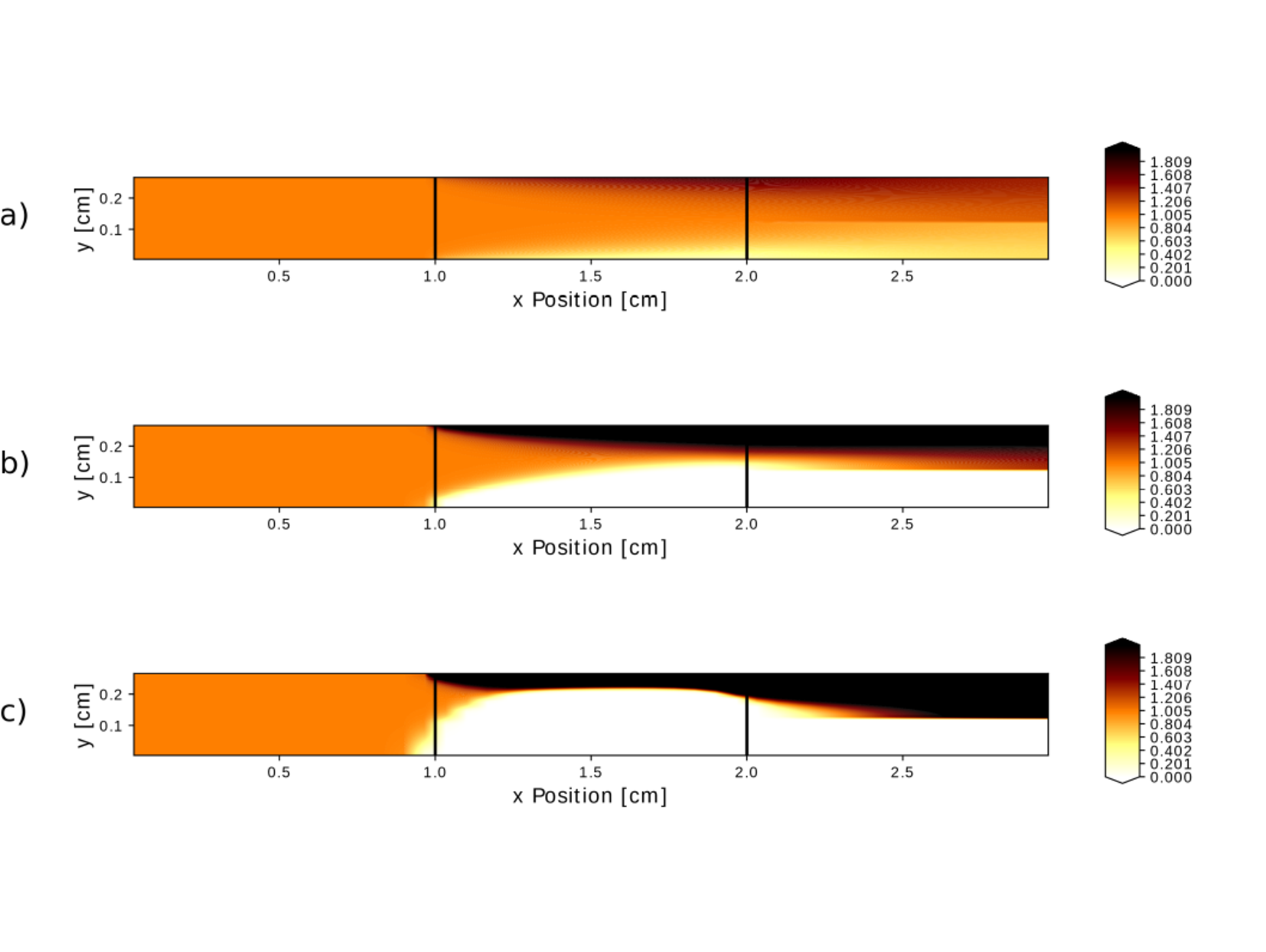}
\caption{Anion concentration profiles for an applied current (per unit width) of a) 15 mA/m, b) 69 mA/m, and c) 204 mA/m, a flow velocity corresponding to an experimental flow rate of 76 $\mu$L/min, and an inlet concentration of 10 mM NaCl.}
\label{SimConc}
\end{center}
\end{figure}

\begin{figure}
\begin{center}
\includegraphics[width=15cm,keepaspectratio,trim=0 90 0 80,clip]{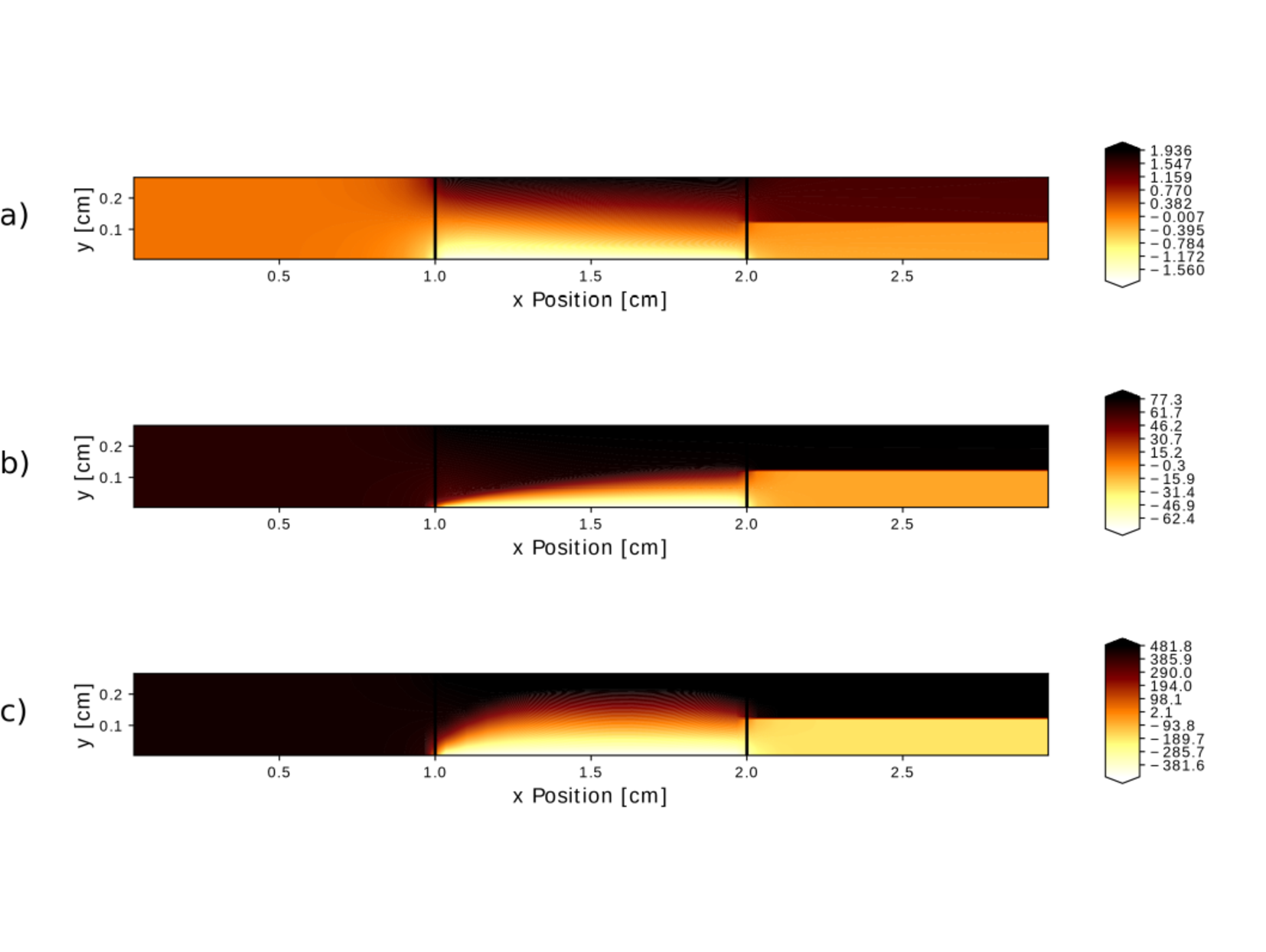}
\caption{Potential profiles for an applied current (per unit width) of a) 15 mA/m, b) 69 mA/m, and c) 204 mA/m, a flow velocity corresponding to an experimental flow rate of 76 $\mu$L/min, and an inlet concentration of 10 mM NaCl.}
\label{SimPot}
\end{center}
\end{figure}

\begin{figure}
\begin{center}
\includegraphics[width=15cm,keepaspectratio,trim=0 90 0 80,clip]{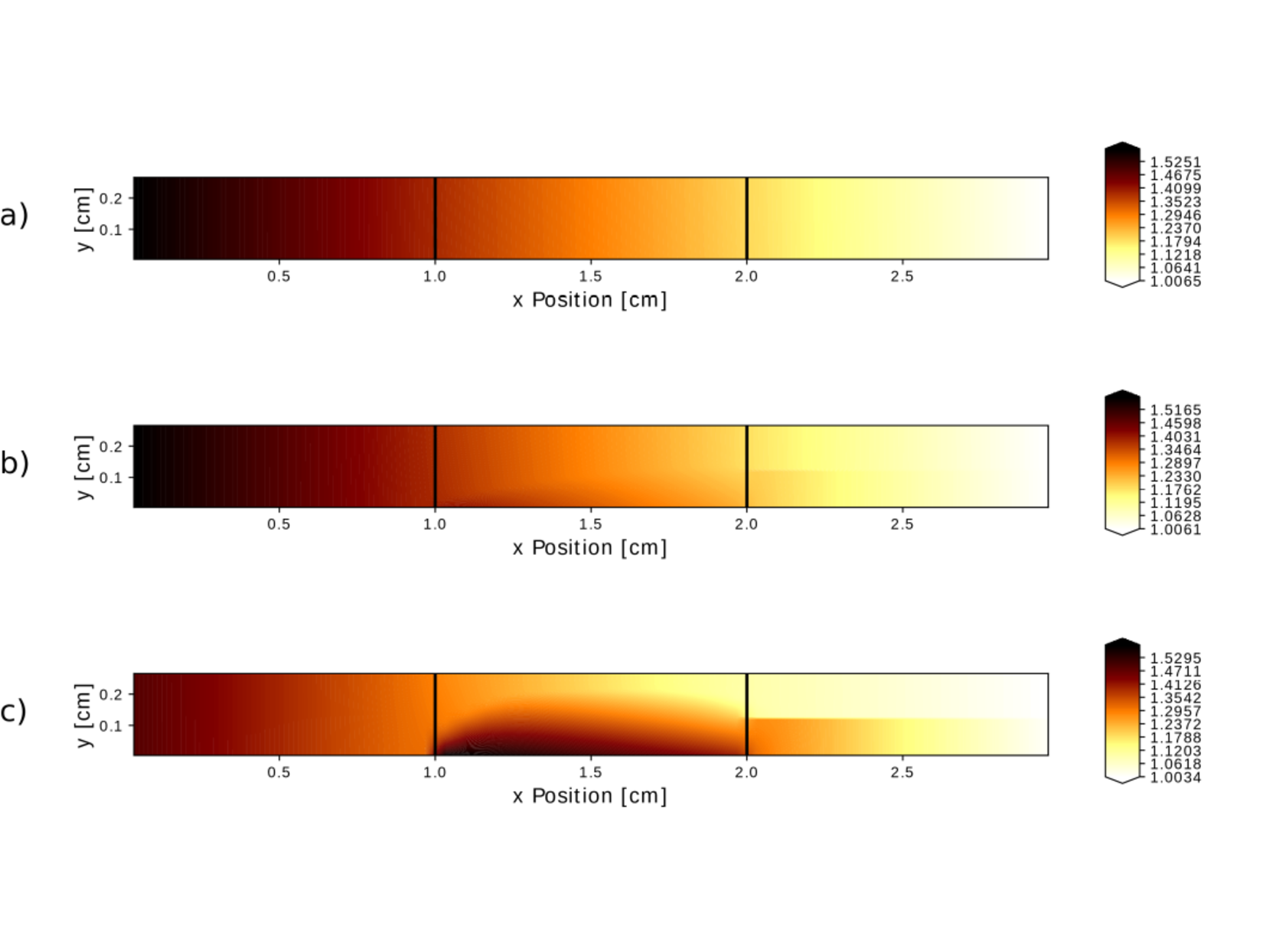}
\caption{Pressure profiles for an applied current (per unit width) of a) 15 mA/m, b) 69 mA/m, and c) 204 mA/m, a flow velocity corresponding to an experimental flow rate of 76 $\mu$L/min, and an inlet concentration of 10 mM NaCl.}
\label{SimP}
\end{center}
\end{figure}

These equations were implemented using the finite volumes method. Simulation results for concentration, potential, and pressure are shown in Figures \ref{SimConc} -- \ref{SimP} for various applied currents. Here we used 33 volumes in the $y$-direction and 45 volumes in the $x$-direction for simulation, which was large enough to produce little error while managing computation time. We studied several binary salt solutions, and the diffusion coefficients used for each ion were $1.334 \times 10^{-9}$ m$^2$/s for Na$^+$, $1.957 \times 10^{-9}$ m$^2$/s for K$^+$, $2.032 \times 10^{-9}$ m$^2$/s for Cl$^-$, $1.90 \times 10^{-9}$ m$^2$/s for NO$_3^-$, and $1.065 \times 10^{-9}$ m$^2$/s for SO$_4^{2-}$ \cite{samson_calculation_2003, cussler_diffusion_2009}.


We can see in these figures that the simulation gives us largely expected results. At very low currents, we see a boundary layer form. As we increase the applied current, such that the residence time is long enough to reach Sand's time, we start observing a shock forming. In the case of the current parameters chosen, we observe the shock reaching the splitter at approximately 69 mA/m, as can be seen in Figure \ref{SimConc}b. At currents above this current, we are doing more work than we need to do, since the shock expands well past the splitter and hence we are wasting energy. In terms of the electrical potential, we see that as soon as the shock forms, the vast majority of the potential drop occurs across the shock region, which corroborates the assumption that was made in the boundary layer model of the system. Lastly, the pressure profiles show that with increasing current, a significant pressure gradient in the $y$-direction starts to develop, which is consistent with the observation that electro-osmotic flow towards the cathode is opposed by pressure-driven flow towards the anode.

\section{Comparison with Experimental Data}

In this part, we consider how the desalination and water recovery predicted by the boundary layer model and the simulation compare to experiment. In the following figures, we scale the current by $z_+ s_+ c_0 e Q$, where $s_+ c_0$ is the inlet concentration of cations and $Q$ is the flow rate, motivated by the experimental data collapse demonstrated in Ref.~\cite{Schlumpberger_Scalable_2015}. This universal scaling will be justified below. 

\begin{figure}
\begin{center}
\includegraphics[width=10cm,keepaspectratio]{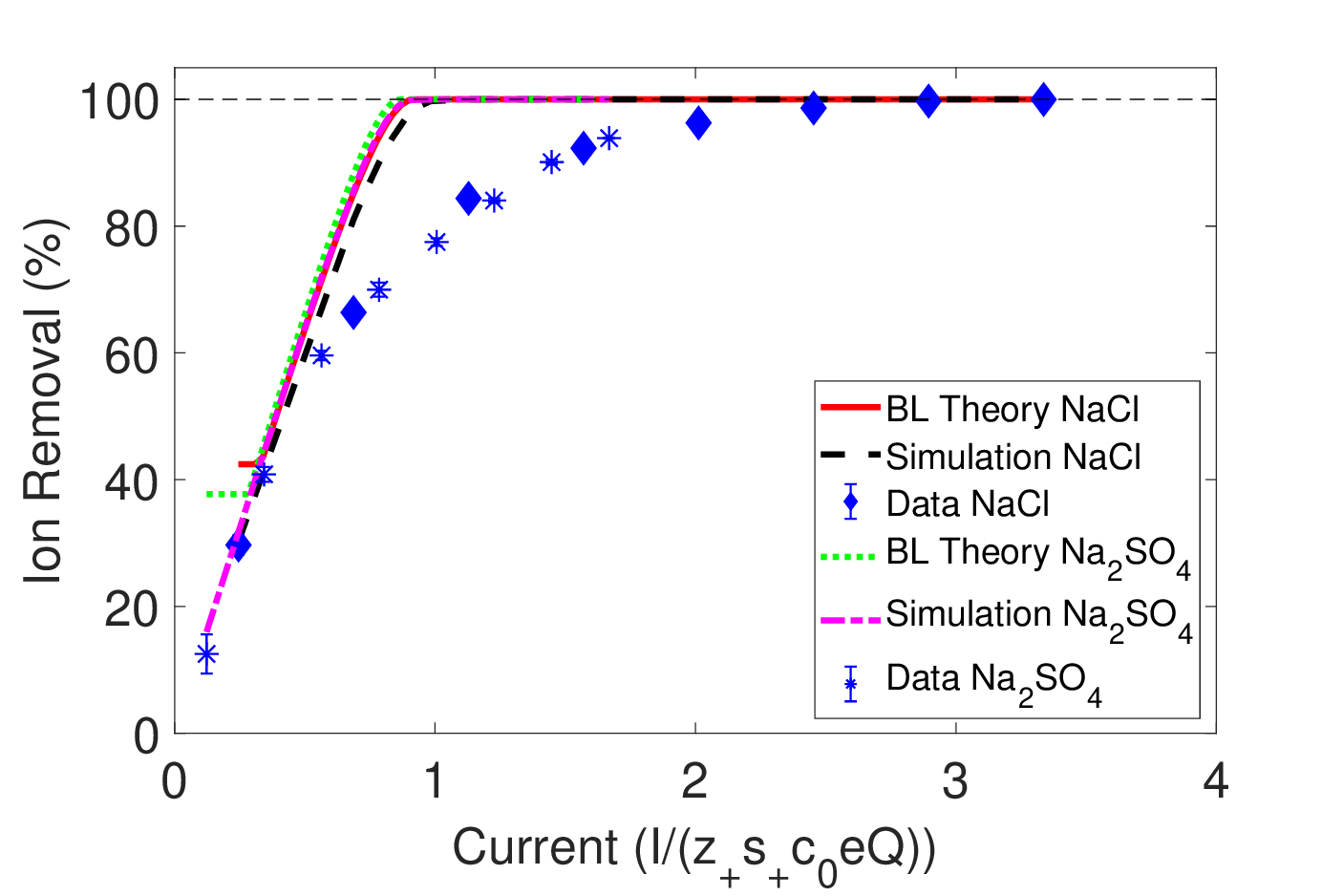}
\caption{Comparison of the ion removal percentage of the boundary layer theory, the simulation, and the experiments for 10 mM NaCl and 10 mM Na$_2$SO$_4$ at conditions corresponding to a flow rate of 76 $\mu$L/min. The simulation appears to predict the experimental values better than the boundary layer theory.}
\label{SimTheory}
\end{center}
\end{figure}

Figure \ref{SimTheory} shows the comparison of desalination efficacy between the prediction by the boundary layer theory, the prediction via simulation, and the measured experimental data for 10 mM NaCl and 10 mM Na$_2$SO$_4$ at conditions corresponding to a flow rate of $Q = 76\ \mu$L/min. At low dimensionless current (less than 0.5), the data appear to indicate that  the simulation is capable of predicting the experimental data somewhat better and the boundary layer theory over-estimates the amount of desalination. 
This limitation is due to the assumption in the boundary layer model that the concentration at the cathode-side cation-exchange membrane is zero; which is not necessarily true at low currents. 
At higher currents, the boundary layer theory and simulation yield close results, and both grossly over-predict the experimental data of desalination.
This discrepancy might be due to the simplistic treatment of electro-osmotic flow and ignoring of H$^+$ transport in the simulation. 

\begin{figure}
\begin{center}
\includegraphics[width=10cm,keepaspectratio]{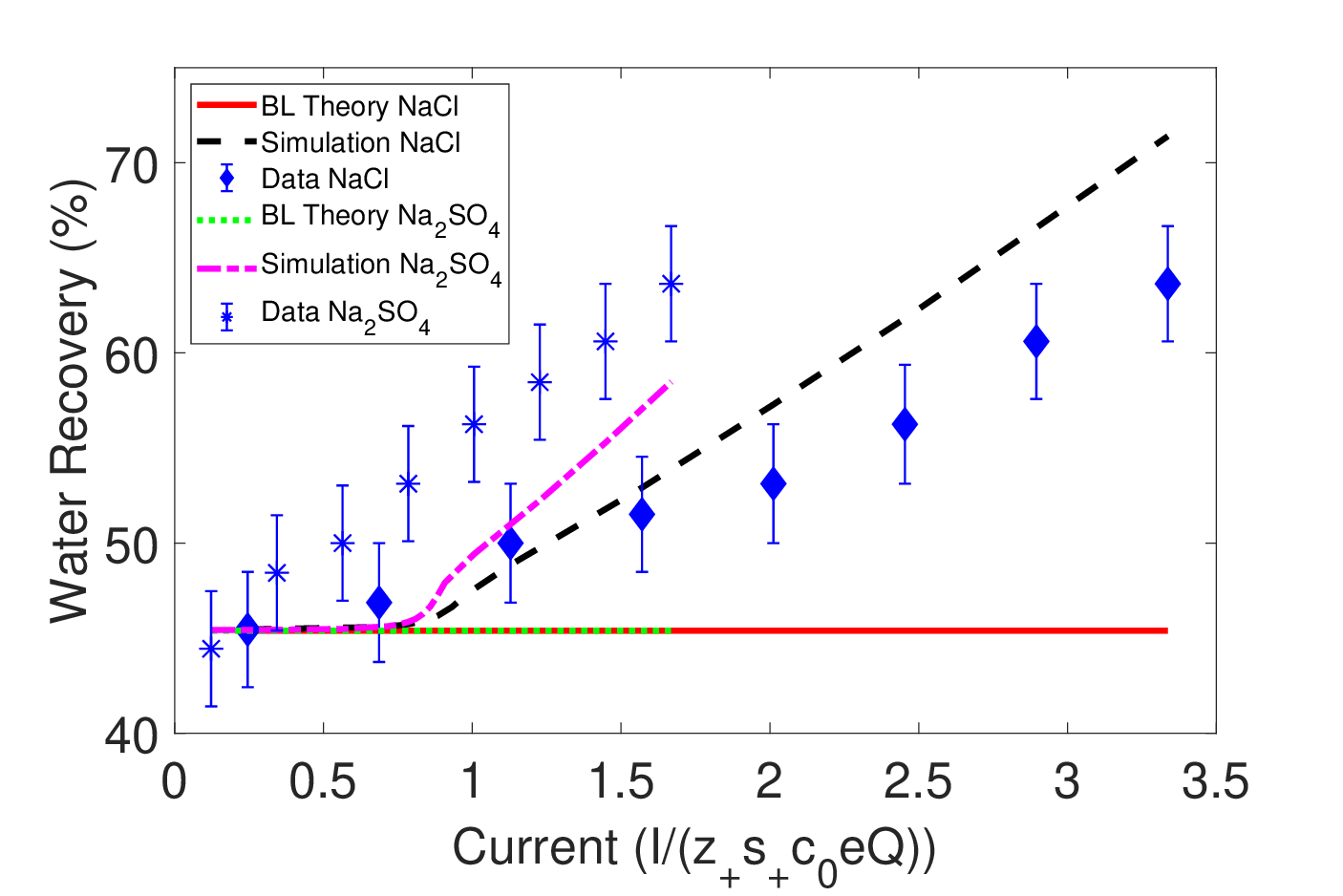}
\caption{Comparison of the water recovery of the boundary layer theory, the simulation, and the experiments for 10 mM NaCl  and 10 mM Na$_2$SO$_4$ at conditions corresponding to a flow rate of 76 $\mu$L/min. The simulation appears to predict the experimental values much better than the boundary layer theory. In fact, the boundary layer theory is not capable of predicting any change in water recovery.}
\label{SimTheoryRecov}
\end{center}
\end{figure}

Furthermore, Figure \ref{SimTheoryRecov} shows the comparison of the water recovery predicted by both theories and the experimental data. As we can see in this figure, the boundary layer model is unable to predict any change in water recovery, whereas the simulation is able to predict the change in water recovery, albeit not always the correct magnitude of the change. The prediction by the simulation is remarkably close to the experimental data for NaCl but deviates substantially for the more complex electrolyte Na$_2$SO$_4$. Also, it is interesting to note that water recovery in the model only starts increasing significantly once the current at which electro-osmotic flow becomes similar in magnitude to pressure-driven flow is reached, whereas the water recovery starts increasing more quickly in experiment. This observation either means that our estimate of the electro-osmotic flow is insufficient (the magnitude is wrong or the way it is incorporated in the model is too simplistic) or that another phenomenon is causing the water recovery to start increasing from the start in experiment.

\begin{figure}
\begin{center}
\includegraphics[width=10cm,keepaspectratio]{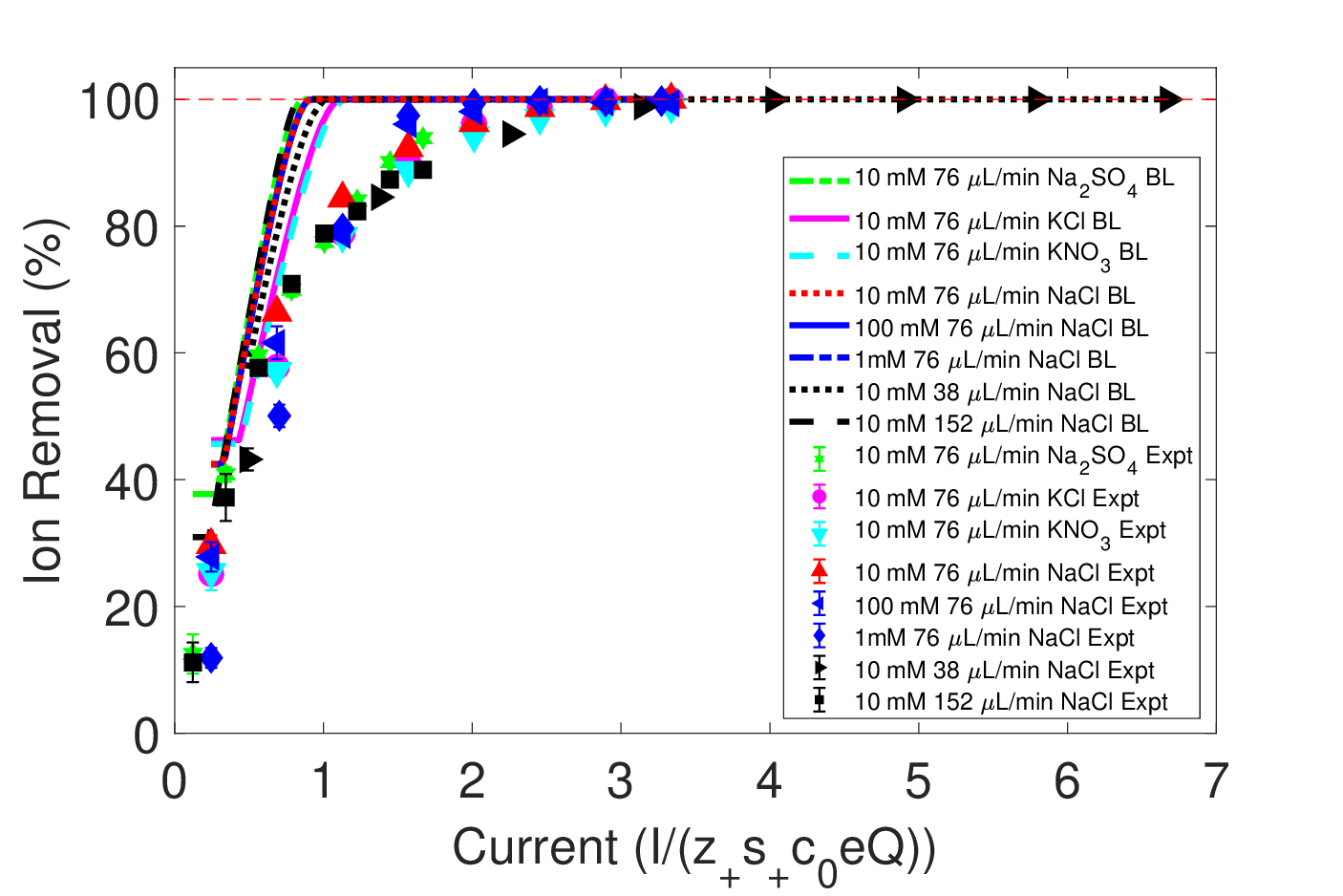}
\caption{Theoretical data predicted by the boundary layer model for various different electrolytes plotted against the same dimensionless current that allowed collapse of the experimental data. Here we can see that these theoretical data do collapse onto one curve using this scaling.}
\label{BLcoll}
\end{center}
\end{figure}

In the experiments of Ref.~\cite{Schlumpberger_Scalable_2015} it was found that by scaling the applied current to the flow rate of total cation charge into the system $z_+s_+c_0 eQ$, the ion removal data could be collapsed onto a master curve for a wide range of symmetric electrolytes and salt concentrations. Thus, we also consider here whether the model similarly collapses with the same scaling. Figure \ref{BLcoll} shows the theoretical data for various electrolytes obtained via the boundary layer model plotted with this linear flow rate scaling. As shown in the figure, the theoretical data do collapse onto a single curve using this scaling. This can also be inferred from the boundary layer equations. Eq. \ref{eq:alpha} indicates $V_{app} \sim \alpha^2 c_{b,0}$ (L'Hôpital's rule), Eq. \ref{BLcurr} indicates that $I \sim V_{app} \sqrt{U_0}/\alpha$, and note that the desalination $\sim \delta_s/H \sim \alpha/\sqrt{U_0}$. Put the above three scalings together, we have desalination $\sim I/c_{b,0}  U_0$, or in the experimental quantities, desalination $\sim I/s_+ c_0 Q$ (we used $c_{b, 0} = -2z_- c_- = 2s_+ c_0$, since $z_+=1$, $s_-=1$ in all the cases).


\begin{figure}
\begin{center}
\includegraphics[width=10cm,keepaspectratio]{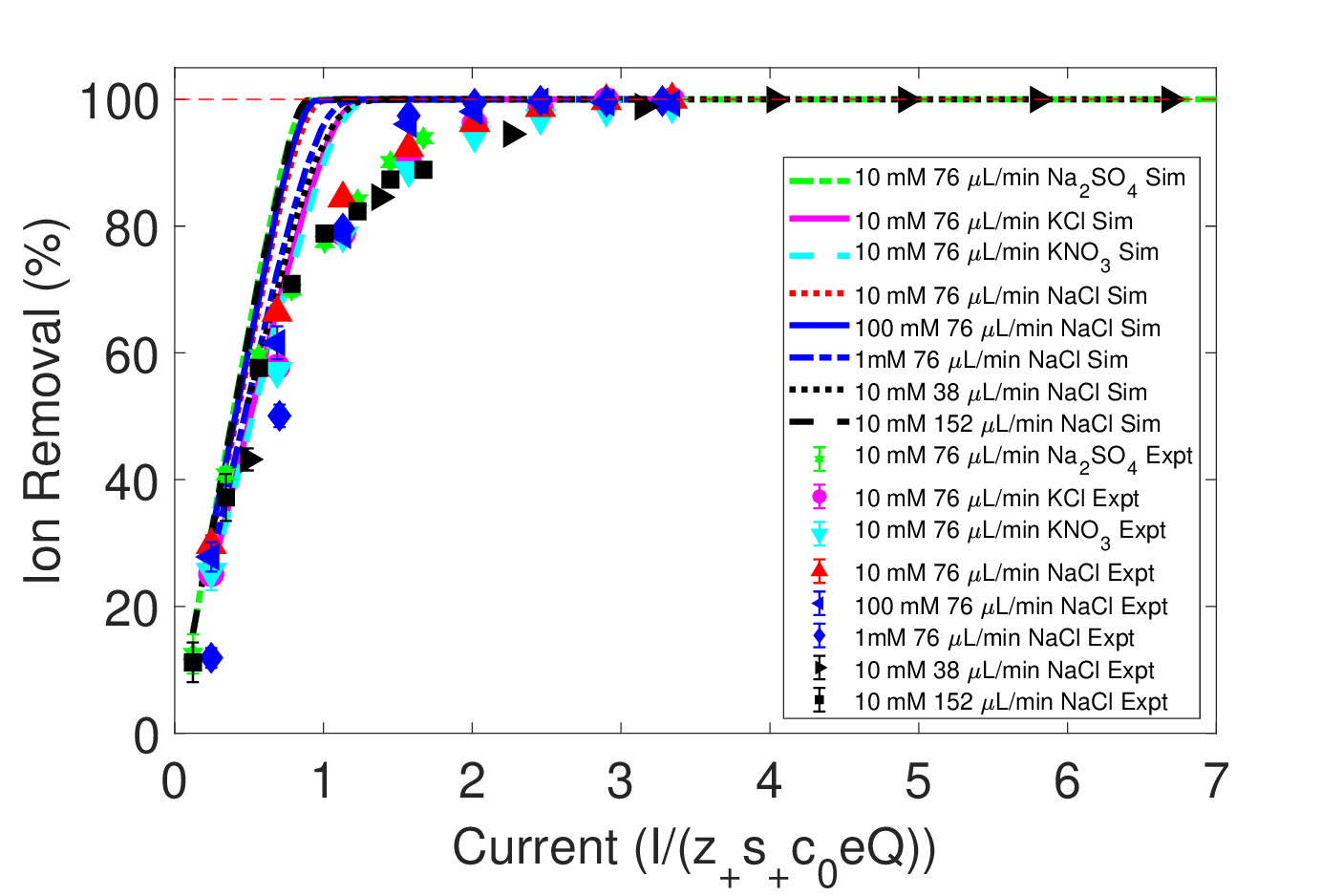}
\caption{Theoretical data predicted by the simulation for various different electrolytes plotted against the same dimensionless current that allowed collapse of the experimental data. Here we can see that, while these theoretical data do collapse onto one curve using this scaling, they do not collapse onto the same curve as the experimental data.}
\label{Simcoll}
\end{center}
\end{figure}

In addition to investigating the collapse for the boundary layer model, we have investigated it for our simulation. Figure \ref{Simcoll} shows the simulation data for various electrolytes at various dimensionless currents, using the scale that allowed collapse of the experimental data. We can clearly see that this scale does lead to collapse of our simulation data, which is consistent with the boundary layer model.

\begin{figure}
\begin{center}
\includegraphics[width=13.9cm,keepaspectratio]{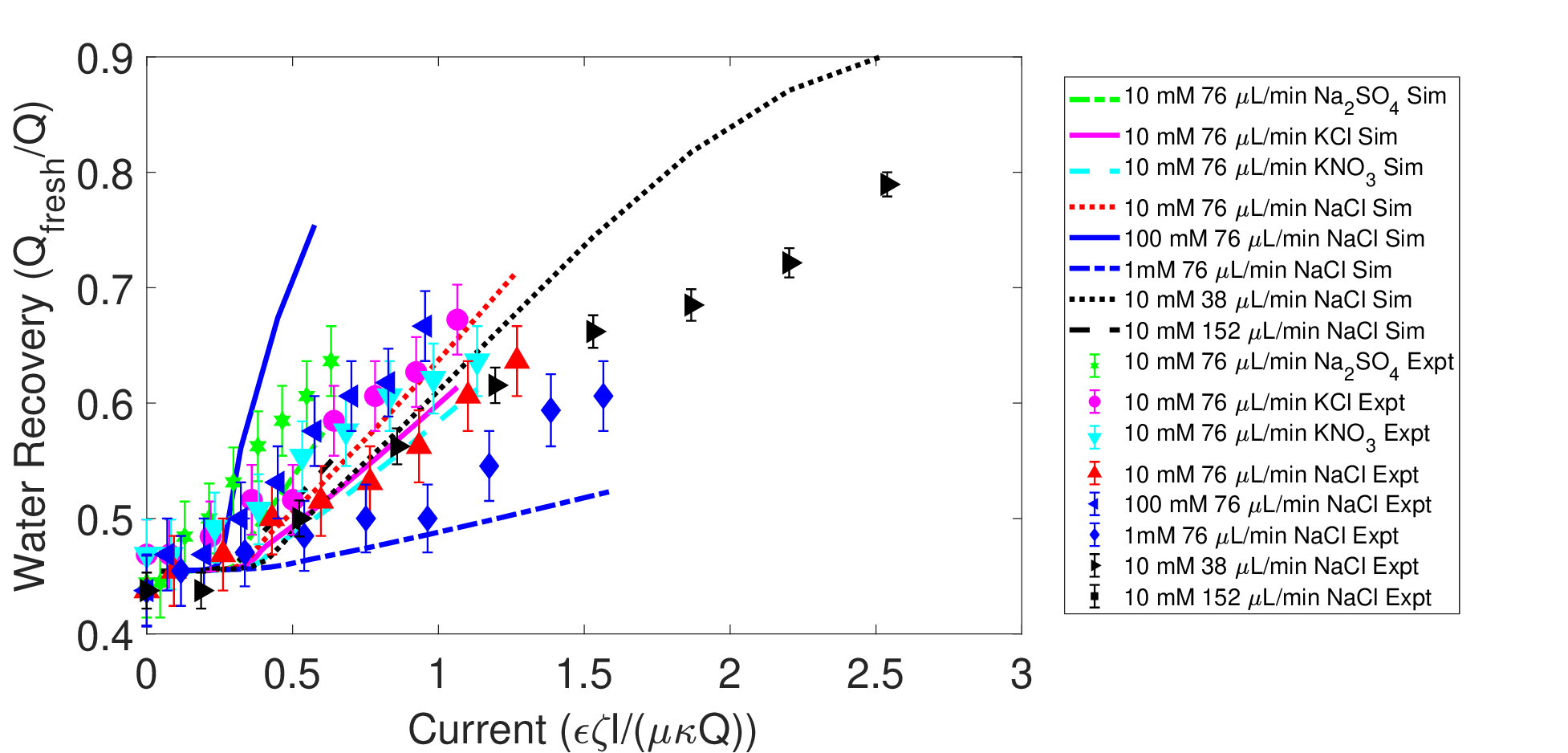}
\caption{Theoretical data for water recovery (lines) predicted by the simulation for various different electrolytes plotted against the same dimensionless current that allowed near-collapse of the experimental data (where $\kappa$ is the bulk electrolyte conductivity). Here we can see that this theoretical data does not appear to collapse onto a single curve nearly as well. Note that we here used the zeta potentials from Table \ref{surfTab}.}
\label{SimCollRecov}
\end{center}
\end{figure}

In contrast to the ability to predict the scale for ion removal, Figure \ref{SimCollRecov} shows that scaling an estimate of the electro-osmotic flow based on the Helmholtz-Smoluchowski formula by the applied flow rate does not yield the correct scaling for water recovery, as the simulation data are not well consistent with the experiments. Some possible causes for this discrepancy could be the simplistic formulation of electro-osmotic flow or the assumption of homogeneous porous media in the model. In fact, there are surely electro-convective vortices at  the pore scale~\cite{dydek_overlimiting_2011,nam2015,bernal2016hydrodynamic}, as well as regions of strong salt depletion and double layer overlap, leading to violation of the assumptions of linear electrokinetic response~\cite{nielsen_concentration_2014} and local electroneutrality~\cite{levy2020breakdown}. The distribution of pore size in the porous media may lead to eddy dispersion, even for linear electrokinetics \cite{Mirzadeh2020VorticesMedia}, and electrokinetic dispersion in a porous network has been shown to strongly enhanced over-limiting current and deionization, both experimentally~\cite{deng_overlimiting_2013} and theoretically~\cite{Alizadeh2019ImpactMedia}. Such effects would presumably need to be captured in future models.



\section{Conclusion}
In summary, we have developed and analyzed a macroscopic leaky-membrane model for the fundamental problem of a deionization shock in cross flow. We derived analytical boundary layer approximations, which are able to accurately capture the desalination results of full numerical solutions of the model. In addition, the numerical simulation can capture the water recovery increase with current. Both approaches are able to provide significant physical insight into the problem, such as predicting the order of magnitude of the currents necessary to achieve desalination. However, even though the simulation is an improvement over prior models, especially with regard to predicting changes in water recovery, the model still fails to capture quantitatively the shape of the scaling functions observed in experiment. Nevertheless, the boundary-layer approximation and full numerical model can serve as a useful basis for making informed decisions about choosing the right materials in experiment, such as choosing a porous medium with an appropriate surface charge.

Improvements to the simulation model can be made by more rigorously treating the flow field (to include dispersion eddies, diffusio-osmosis, etc.), including multiple-ion transport, and resolving the profiles in the cross-sections of the pores.  The first improvement would likely be useful in more closely matching the model's ion removal predictions with those observed in experiment at moderate currents, because they would more rigorously account for enhanced mixing in the depleted region. The second improvement would allow us to investigate the expected current efficiencies in the system by accounting for the current from water splitting, and to investigate the selective ion removal observed in recent shock electrodialysis experiments \cite{Conforti2019ContinuousElectrodialysisb, Alkhadra2019ContinuousElectrodialysis, Alkhadra2020Small-scaleElectrodialysis}. 

\begin{acknowledgments}
This research was supported by Weatherford International, the MIT Energy Initiative, and the SUTD-MIT Fellowship Program.
\end{acknowledgments}

\bibliography{SimPaper_v2} 

\end{document}